\documentclass{CVM}

\def\method{TexPro\xspace}

\CVMsetup{
type      = {Research/Review Article},
doi       = {CVM.XXXX},
title     = {TexPro: Text-guided PBR Texturing with Procedural Material Modeling},
author    = {Ziqiang Dang$^{1*}$, Wenqi Dong$^{1*}$, Zesong Yang$^{1}$, Bangbang   Yang$^{2}$, Liang Li$^{3}$, Yuewen Ma$^{2}$, and Zhaopeng Cui$^{1}$\cor{}\\
},
runauthor = {Ziqiang Dang et al.},
abstract  = {
  In this paper, we present \method, a novel method for high-fidelity material generation for input 3D meshes given text prompts. Unlike existing text-conditioned texture generation methods that typically generate RGB textures with baked lighting, \method is able to produce diverse texture maps via procedural material modeling, which enables physically-based rendering, relighting, and additional benefits inherent to procedural materials.
  Specifically, we first generate multi-view reference images given the input textual prompt by employing the latest text-to-image model. 
  We then derive texture maps through rendering-based optimization with recent differentiable procedural materials. 
  To this end, we design several techniques to handle the misalignment between the generated multi-view images and 3D meshes, and introduce a novel material agent that enhances material classification and matching by exploring both part-level understanding and object-aware material reasoning.   Experiments demonstrate the superiority of the proposed method over existing SOTAs, and its capability of relighting.
},
keywords  = {3D asset creation, text-guided texturing, procedural materials, MLLM},
copyright = {The Author(s)},
}

\begin{document}
\def\etc{\emph{etc\dot}\xspace}
\def\eg{\emph{e.g.}\xspace} 
\def\Eg{\emph{E.g.}\xspace}
\def\ie{\emph{i.e.}\xspace} 
\def\Ie{\emph{I.e.}\xspace}
\def\etal{\emph{et al.}\xspace}
\newcommand{\figref}[1]{{\ref{#1}}}
\maketitle

\enlargethispage{-3pt}
\begin{figure}[b] \vskip -4mm
\small\renewcommand\arraystretch{1.3}
\begin{tabular}{p{80.5mm}} \toprule\\ \end{tabular}
\vskip -4.5mm \noindent \setlength{\tabcolsep}{1pt}
\begin{tabular}{p{3.5mm}p{80mm}}
$1\quad $ & Zhejiang University, hangzhou, 310058, China. E-mail: Ziqiang Dang, ZiqDang@zju.edu.cn; Wenqi Dong, dongwenqi@zju.edu.cn; Zesong Yang, zesongyang0@zju.edu.cn; Zhaopeng Cui, zhpcui@zju.edu.cn\cor{}.\\
$2\quad $ & Pico, ByteDance Inc, shanghai, 200235, China. E-mail: Bangbang Yang, ybbbbt@gmail.com; Yuewen Ma, mayuewen@bytedance.com.\\
$3\quad $ & Communication University of Zhejiang, Hangzhou, 310018, China. E-mail: Liang Li, liliang@cuz.edu.cn.\\
$*\quad $ & Authors contributed equally.\\
% &{\textcolor{blue} {(If the authors are from the same affiliation, then the same superscript should be marked on each author's name. Please provide each author's official email address)}}\\
&\hspace{-5mm} Manuscript received: 2025-01-01; accepted: 2025-01-01\vspace{-2mm}
\end{tabular} \vspace {-3mm}
\end{figure}

\section{Introduction}
\label{sec:intro}

% Introduce the mesh texturing
3D mesh texturing is a critical process in the realm of 3D modeling and visualization, which holds significant importance across various applications such as AR/VR, gaming, filmmaking, \emph{etc}.
By transforming a basic geometric shape into a visually detailed and realistic object, mesh texturing not only improves the aesthetic appeal of the model but also greatly enhances immersion and realism in digital environments. 
To achieve photorealistic rendering that responds accurately to varied lighting conditions, %textures must go beyond basic color information to 
we need to generate object material that includes a set of texture maps for diffuse and specular reflections, glossiness, surface roughness, \emph{etc}. 
Thus, traditional mesh texturing typically involves substantial manual effort. 
Recent advances in 2D image generation have spurred the development of automated mesh texturing methods~\cite{chen2023text2tex,richardson2023texture}, but they normally generate RGB textures with baked lighting, which limits their applicability in downstream tasks. 

\begin{figure*}[!t]
    \centering
    \includegraphics[width=0.96\linewidth]{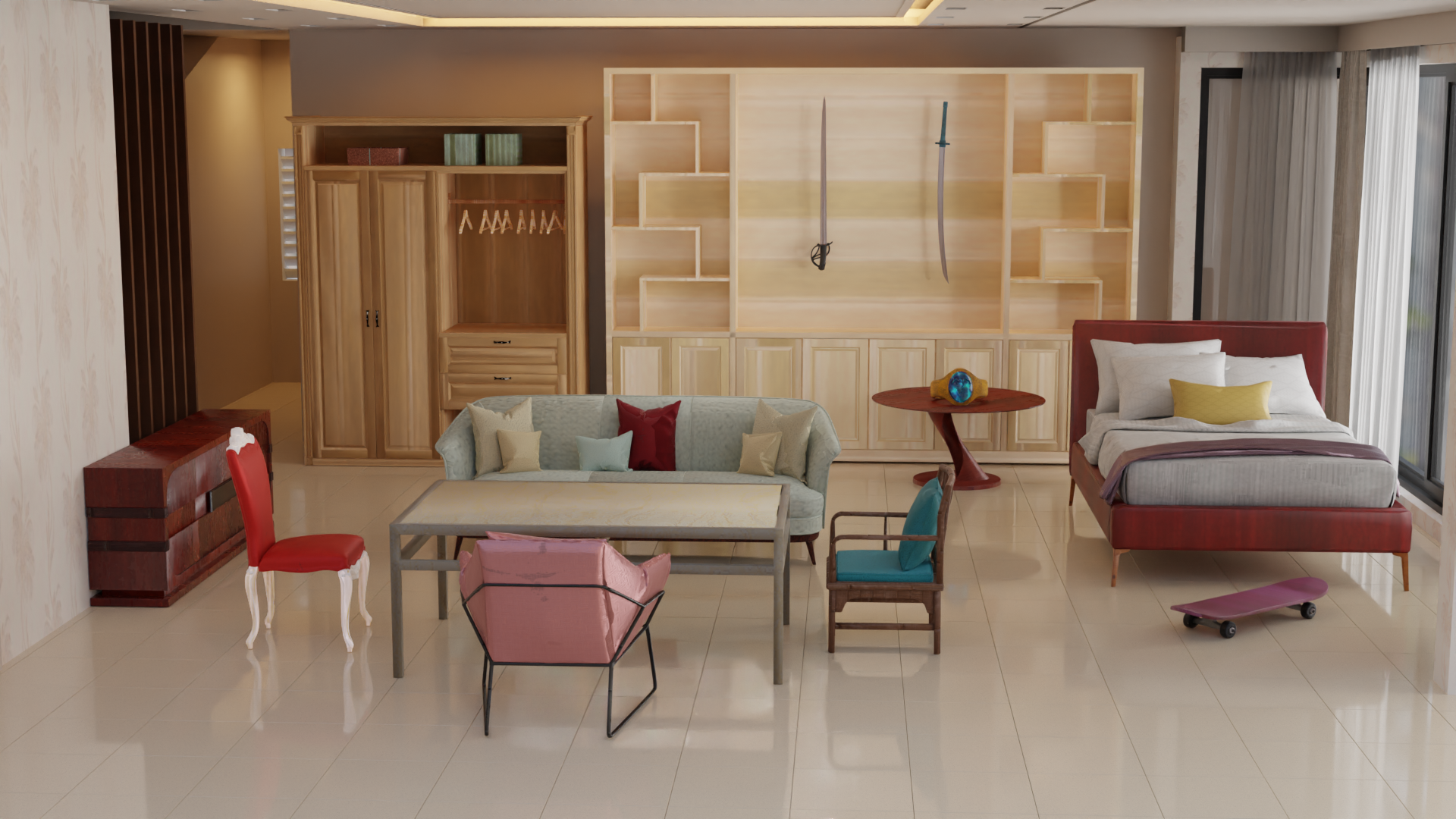}
    % \vspace{-0.1cm}
    \caption{{Texturing Results.} We introduce \method, a text-driven PBR material generation method for photorealistic and relightable rendering.
    }
    \label{fig:Teaser}
    \vspace{-0.1cm}
\end{figure*}

In this paper, we propose a novel method for producing high-quality materials for 3D meshes based on user-provided text prompts as shown in Fig.~\ref{fig:Teaser}. Unlike previous methods that directly generate pre-baked 2D texture maps, 
our approach is able to produce diverse texture maps via procedural material modeling, which enables physically-based rendering, relighting, and additional benefits inherent to procedural materials, such as resolution independence (\ie, the ability to producing textures at any desired resolution), re-editability, %the ability to change dynamically at runtime, 
efficient storage, and so on.
The proposed system could simplify the workflow for
amateur artists, eliminating the need to spend time on tedious
parameter tuning for procedural materials. Moreover, the system’s output can serve as useful initialization, allowing artists to further adjust the parameters of nodes in the procedural material graph to achieve the desired detailed effects without starting from scratch.

Specifically, our approach first generates multi-view reference images given the input textual prompt by employing the latest text-to-image models~\cite{rombach2022high,controlnet}, and then obtains diverse texture maps through rendering-based optimization with recent differentiable procedural materials~\cite{shi2020match,li2023end}. After the optimization, we can essentially obtain procedural materials for the mesh, thus supporting downstream applications.
However, it is non-trivial to design such a system.

First,  despite using additional rendered depth or normal maps as conditions~\cite{controlnet}, the generated images from the text-to-image model (\eg, Stable Diffusion~\cite{rombach2022high}) often lack accurate alignment with the meshes and may not maintain consistency across different views. 
To handle these misalignment challenges, as shown in Fig.~\ref{fig:pipeline}, we propose to exploit an improved segmentation method, Matcher~\cite{liu2024matcher}, %based on SAM~\cite{kirillov2023segment} 
to extract the masks of corresponding objects in generated images, which will be further smoothed and %intersected 
integrated with the rendered masks to obtain the final aligned masks.
Additionally, we design an adaptive camera sampling method 
that ensures each material part exceeds a specified pixel count threshold and is assigned a unique material to address the inconsistency across views.

Second, 
an initial estimation of material properties is requisite for differentiable procedural material rendering.
However, the presence of baked lighting in generated images and the confusion of different materials with similar color distributions challenge initial material classification and matching  for classical methods~\cite{photoshape2018,hu2022photo}.
To overcome these challenges, 
we leverage the capabilities of multi-modal large language models (MLLMs) to create a novel material agent. Given %\ziqiang{\sout{the reference patches} 
the masked prompt image, the material agent is able to provide the possible material types in the order of likelihood for  %marked 
the corresponding object via part-level understanding and object-aware material reasoning,  which further facilitates the rendering-based optimization with procedural materials.

Additionally, 
to better recover the material properties, we also consider the environmental lighting during optimization. To make the optimization tractable, we adopt a plain lighting setup to fit the lighting environment in the image. Given the object mesh, our method adaptively creates the floor and four walls in the scene of the differentiable renderer, and adaptively places initial area lights on each wall and above the object. Then we optimize the RGB intensity of these lights together with procedural materials. 

Our contributions can be summarized as follows:
\begin{itemize}[itemsep=0pt, topsep=0pt]
\item We propose a novel framework that is able to create high-fidelity textures for input meshes given text prompts and enables physically-based rendering (PBR) and relighting.
\item In order to handle the misalignment between the generated multi-view images and 3D meshes, 
we propose to combine the rendered mask with an advanced segmentation method, Matcher~\cite{liu2024matcher}, 
to achieve accurately aligned part-level masks, and design an adaptive camera sampling strategy %is designed 
to deal with inconsistency across views.
\item We present a novel material agent for robust material classification and matching by exploring both part-level understanding and object-aware material reasoning.
\item The experiments show that the proposed method outperforms the existing state of the art, and maintains the capabilities of relighting.
\end{itemize}

\section{Related Work}
\subsection{Texture Generation}
Texture maps are essential for 3D geometric models. They involve mapping pre-defined 2D planar images onto 3D models, endowing 3D models with vital color information. Traditional texture creation involves cumbersome manual drawing, assembling repetitive patterns, or stitching multi-view images~\cite{waechter2014let,bi2017patch}.
With the development of increasingly high-quality 3D datasets and advances in text-to-image generative techniques~\cite{ramesh2022hierarchical,rombach2022high,controlnet}, learning-based approaches have been proposed to generate high-quality textures~\cite{khalid2022clipmesh,Michel_2022_CVPR,chen2023text2tex,richardson2023texture,yu2023texture}. 
Text2Tex~\cite{chen2023text2tex} incorporates a depth-aware image inpainting model to synthesize high-resolution partial textures progressively from multiple viewpoints, which has also been widely employed in text-to-scene tasks~\cite{fridman2024scenescape,hollein2023text2room,zhang2024text2nerf}.
Similarly, TEXTure~\cite{richardson2023texture} also applies an iterative scheme to paint 3D models, while presenting a trimap representation and a novel elaborate diffusion sampling process to tackle inconsistency issues.
Although existing methods are capable of generating high-quality textures, they are limited to generating only diffuse maps with baked lighting, lacking support for relighting. Additionally, the generated textures often suffer from the Janus (multi-faced) problem~\cite{armandpour2023re} and lack realism. 

\subsection{Procedural Material Modeling and Capture}
Procedural material has been the standard method of material modeling in industry. Materials are represented as node graphs denoted with simple image processing operations which are combined to produce real-world spatially varying BRDF~\cite{burley2012physically} material maps. 
Given user-specified images or text prompts, many approaches~\cite{hu2019novel,shi2020match,hu2022node,guerrero2022matformer,li2023end} have been proposed to recover the parameters of a predefined procedural material graph in agreement with the input. Moreover, Hu~\etal~\cite{hu2023generating} moved beyond parameter regression, utilizing the diffusion model to generate graph structures from text or image conditions.
It is worth noting that MATch~\cite{shi2020match} proposes a differentiable procedural material method that translates procedural graphs into differentiable node graphs, enabling single-shot high-quality procedural material capture. 

\subsection{Differentiable Rendering}
Differentiable rendering has consistently remained a significant research topic in computer graphics and vision. The gradient in rendering is required with respect to numerous factors, and several specifically differentiable renderers have been developed and widely used in the community, like Redner~\cite{Li:2018:DMC}, Mitsuba 2/3~\cite{nimier2019mitsuba,Mitsuba3}, Nvdiffrast~\cite{Laine2020diffrast} and PSDR-CUDA~\cite{Zhang:2020:PSDR}, the aim being to address challenges arising from the non-differentiable terms in the rendering integral.
We make use of differentiable rendering to backpropagate the gradient from pixel space to the parameters of procedural materials, thereby optimizing the materials via gradient descent.

\section{Method}
\subsection{Aims and Approach}
Most 3D meshes in online resource repositories like BlenderKit or Sketchfab, as well as meshes from existing 3D datasets~\cite{fu20213d,li20223d_compat,slim_3dcompatplus_2023,deitke2023objaverse,deitke2023objaversexl}, have been segmented into different material parts by their authors during creation, \eg, a chair consists of the frame, backrest and cushion.
Moreover, part-level segmentation of a complete mesh can be achieved through shape analysis methods~\cite{hanocka2019meshcnn,lahav2020meshwalker,yin2023sai3d}. 
Thus, given textual prompts, we aim to generate appropriate materials for each material part of the input 3D mesh model and achieve photorealistic and relightable texturing.

\begin{figure*}[!t]
    \centering
    \includegraphics[width=1\linewidth]{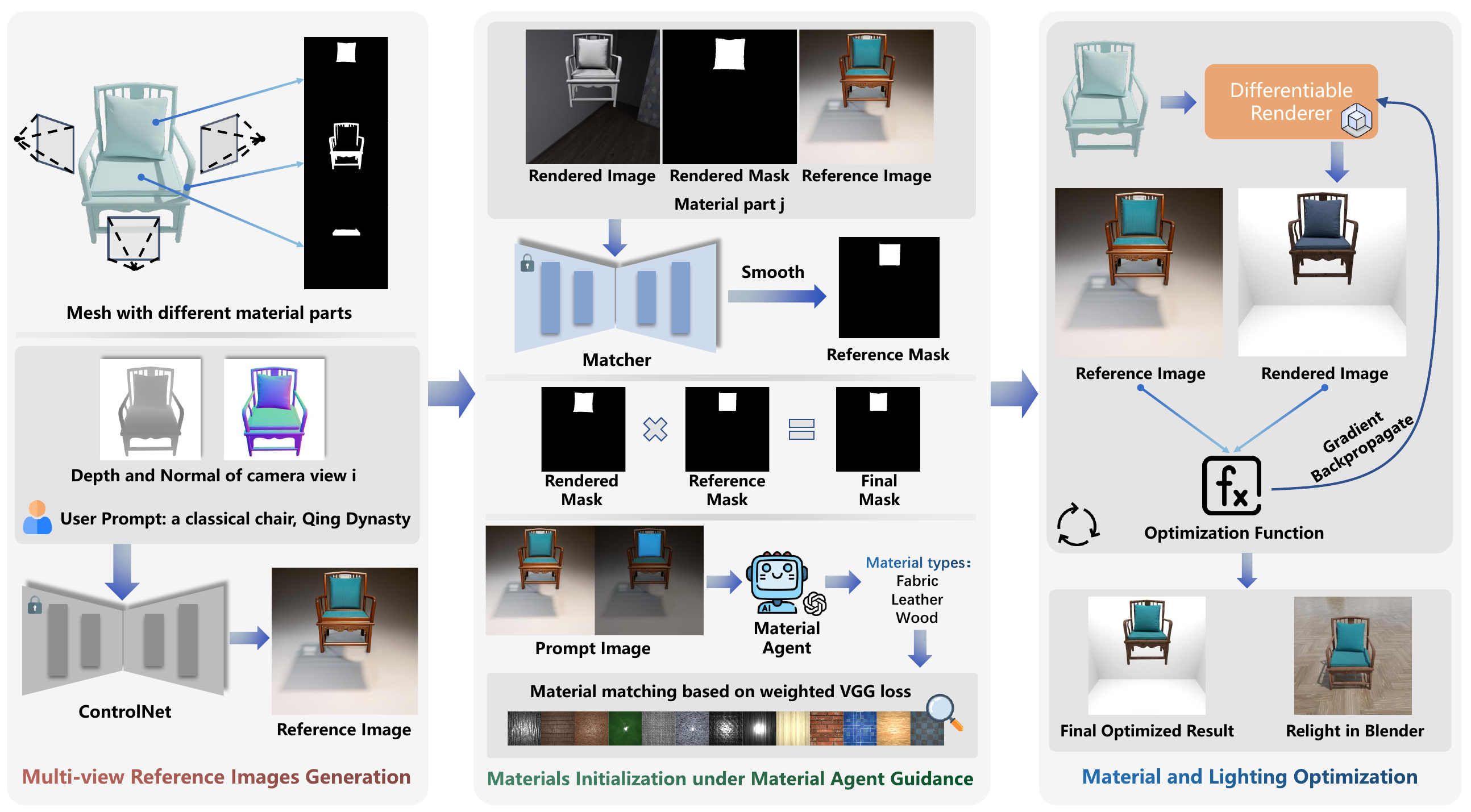}
    % \vspace{-0.7cm}
    \caption{Overview of \method.}
    \label{fig:pipeline}
    % \vspace{-0.2cm}
\end{figure*}
As shown in Fig.~\ref{fig:pipeline}, 
our method textures the mesh with PBR materials through three steps: multi-view reference images generation, materials initialization  %selection 
under material agent guidance, and materials optimization using differentiable rendering. 
In the rest of this section, we first introduce procedural materials and a differentiable version \textit{DiffMat} in Sec.~\ref{ssub:preliminary}.
We then provide detailed descriptions of these three steps in Sec.~\ref{ssub:reference}, Sec.~\ref{ssub:agent} and Sec.~\ref{ssub:optimization}.

\subsection{Preliminaries}
\label{ssub:preliminary}
\noindent\textbf{Procedural materials.}
Procedural material (\eg, Substance materials~\cite{SubstanceDes}) is a standard way of modeling spatially-varying materials in the 3D design industry, which generates texture maps algorithmically.
Different from the static, pre-drawn or pre-baked texture maps, procedural materials are represented as directed acyclic node graphs, primarily consisting of two types of nodes: generators and filters. Generator nodes create spatial textures from scratch, and filter nodes manipulate input textures using certain image processing operations. Specifically, each node has specific parameters, and changing these parameter values will produce different texture maps, including base color, roughness, normal and other maps. 
Procedural materials provide many beneficial properties, 
such as resolution-independence, re-editability, the ability to be dynamically changed during runtime, and greater efficiency in terms of storage and memory usage.

\noindent\textbf{DiffMat.} 
Shi \etal developed the \textit{DiffMat} library~\cite{shi2020match}, skillfully utilizing convolutional neural networks to implement the functionality of most filter nodes, thereby making procedural materials differentiable.
Recently, \textit{DiffMatV2}~\cite{li2023end} was proposed to implement differentiable generator nodes and expand the scope of optimizable parameters of filter nodes. In this work, we utilize \textit{DiffMatV2} to optimize the parameters $\theta$ of the node graphs.

\subsection{Multi-view Reference Images Generation}
\label{ssub:reference}

\noindent\textbf{Camera views selection.}
Since the input mesh is assumed to be segmented beforehand, %already segmented, 
the mask $M_R$ for each material part 
in a certain camera view 
can be rendered directly through the renderer. We first adaptively select some camera views to ensure that the number of pixels for each material part exceeds a certain threshold. Note that the front view has to be selected because the reference image of this view generated by Stable Diffusion~\cite{rombach2022high} is more realistic. The number of selected views is unconstrained.
For more details of the camera selection strategy, please refer to the appendix.

\noindent\textbf{Reference images generation.}
We use the differentiable renderer PSDR-CUDA~\cite{Zhang:2020:PSDR} to render the normal maps and depth maps of the mesh under the selected camera views. Subsequently, we use Stable Diffusion and Controlnet~\cite{controlnet} to generate the reference image of the front view under the condition of the corresponding normal map. For other camera views, we also use the reference image of the front view as an additional condition to generate reference images. Nonetheless, there may still be slight differences between images of different views. 
To deal with this problem, we select only one reference image for each material part according to the number of pixels in each view. Moreover, if the image quality is poor, the rendered depth maps are added as an extra condition.

\subsection{Materials Initialization with Agent Guidance}
\label{ssub:agent}
\begin{figure}[!t]
    \centering
    \includegraphics[width=1\linewidth]{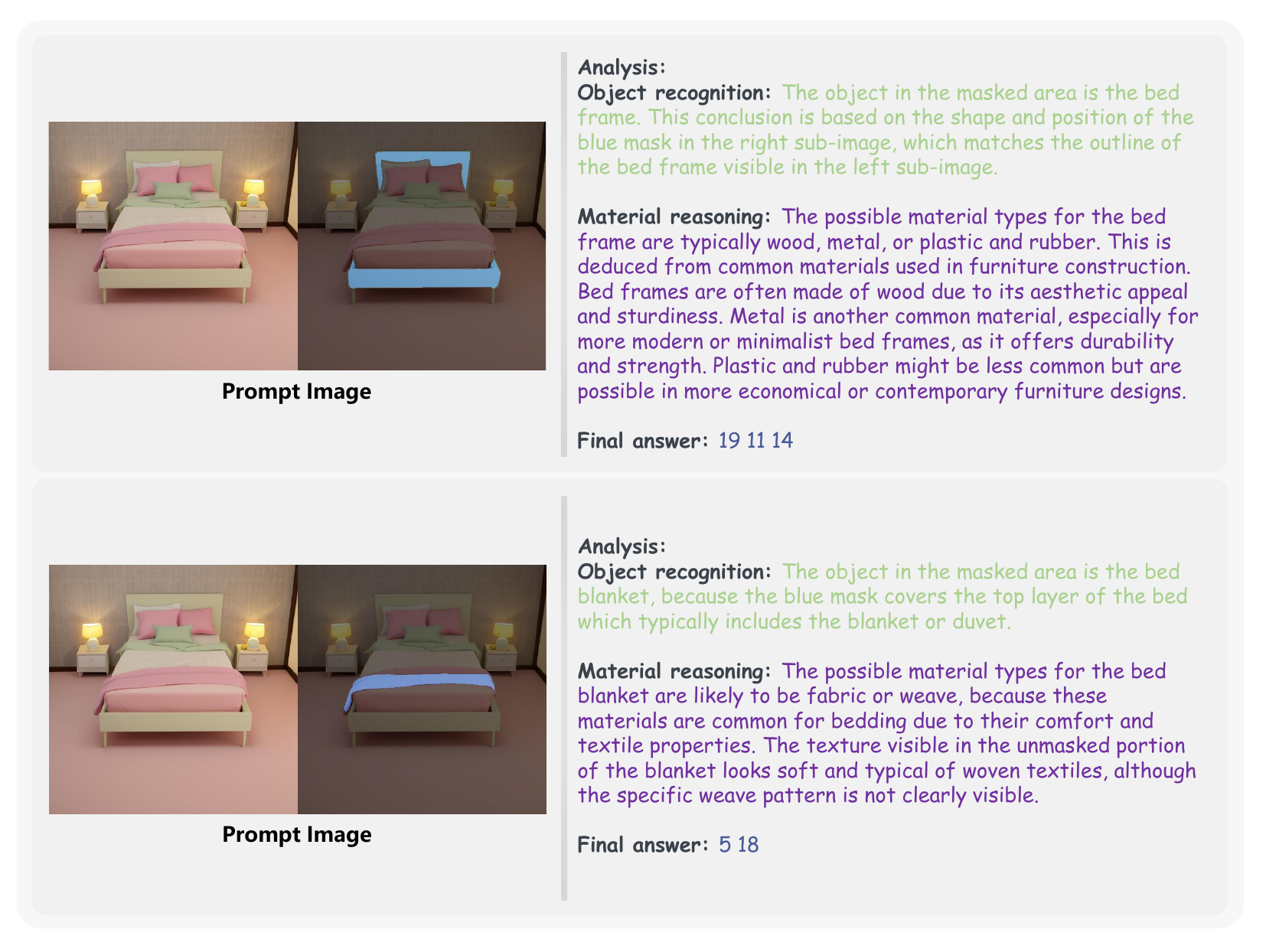}
    % \vspace{-0.2cm}
    \caption{Agent Response Examples. Left: prompt images provided to the material agent. Right: responses returned by the agent.
    }
    \label{fig:response}
\end{figure}

Since an initial estimation of material properties is requisite for differentiable procedural material rendering, we adopt a two-step approach to select the initial procedural material: first by classification, then by feature matching. 
 \begin{figure}[!t]
    \centering
    \includegraphics[width=1\linewidth]{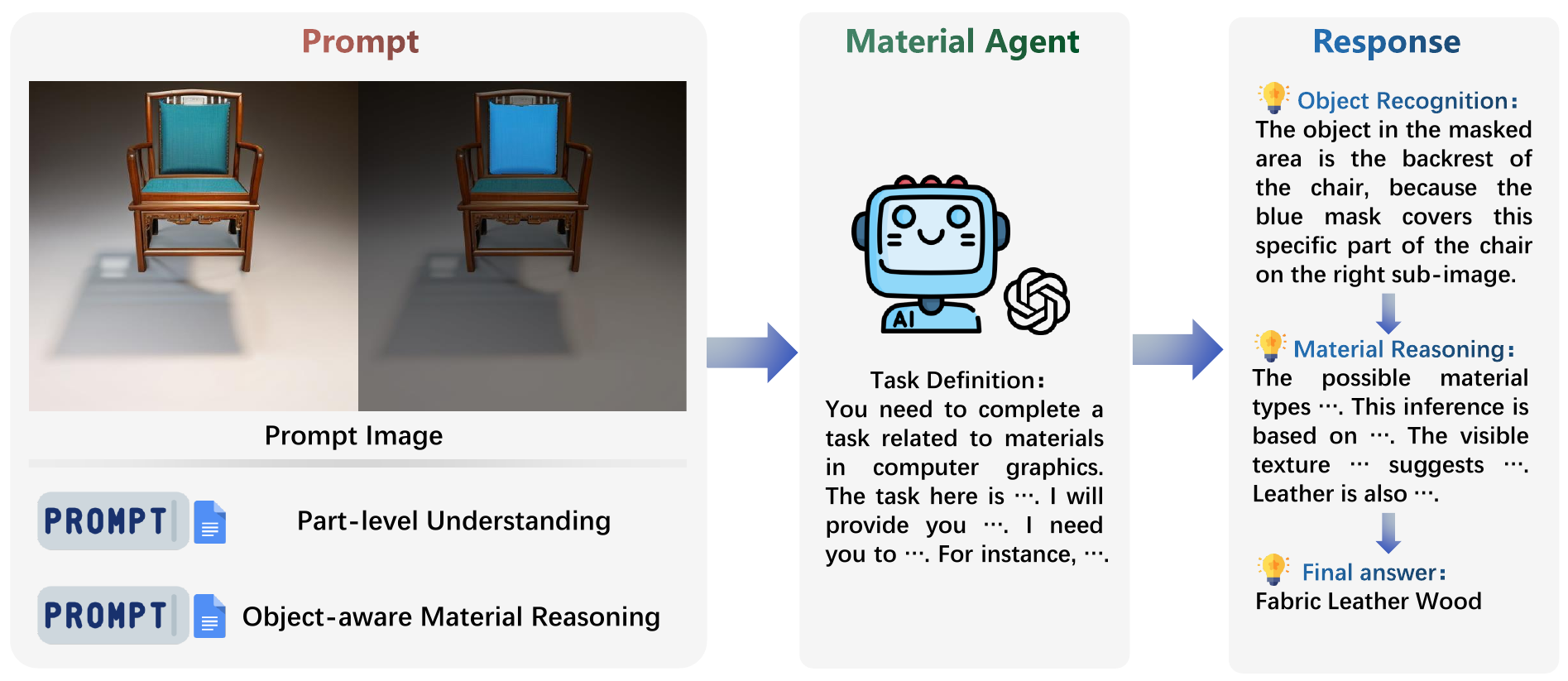}
    % \vspace{-0.6cm}
    \caption{Material Agent. An example of a chair backrest is shown here. Providing both the image prompt and text prompt to our defined material agent allows it to predict all possible material categories in order of likelihood.
    }
    \label{fig:Agent}
\end{figure}
However, we find that classical material classification methods~\cite{photoshape2018,hu2022photo} suffer from the baked lighting in generated images and confuse different materials with similar color distributions. 
Thus, we design a novel material agent using GPT-4V for accurate material initialization via part-level understanding and object-aware material reasoning: for example, pillows should be assigned fabric or leather materials, not metal or wood.
Specifically, as shown in Fig.~\ref{fig:Agent}, given the masked prompt image and the text prompts, the material agent will provide all possible material types in the order of likelihood. 
Additionally, Fig.~\ref{fig:response} presents some examples of agent response and Fig.~\ref{fig:ablation-agent} illustrates the complete classification process and results for a bed mesh.

\noindent\textbf{Mask segmentation and alignment.}
Since the object geometry in the images generated by Stable Diffusion is not well aligned with the rendered images, the mask $M_R$ on the rendered images cannot be directly applied to the reference images. 
Therefore, we use Matcher~\cite{liu2024matcher} to get the mask $M_I$ for the reference images. Specifically, as shown in Fig.~\ref{fig:pipeline}, for camera view $i$, we provide three inputs to Matcher: the rendering of the object textured using white material under specific ambient lighting, the mask $M_R$ of a material part on the rendered image, and the reference image.
Matcher then determines the corresponding mask $M_I$ of the material part on the reference image. Next, we use median filtering, morphological operations, and region area filtering to smooth $M_I$ obtained by Matcher. The final mask $M_{ij}$ of material part $j$ under camera view $i$ is obtained by $M_{ij}=M_R*M_I$. $M_{ij}$ is used on both rendered images and reference images.

\noindent\textbf{Material agent prompt design.} 
For each material part, we blend its mask $M_{ij}$ with the reference image, and then concatenate it horizontally with the reference image to create the prompt image. Given the name of the mesh (\eg, chair), a specific prompt text is derived from our carefully designed prompt template to guide the material agent. Note that the same prompt text is shared across all material parts, instead of using a separate one for each. Our template starts by using the GPT-4V system prompt obtained through reverse prompt engineering to enhance performance. Then, we provide a detailed definition of the task along with an example for in-context learning~\cite{min2022rethinking}. Subsequently, we employ various prompt techniques to construct the prompt for part-level understanding and object-aware material reasoning.
Please refer to the appendix for the complete prompt template.

\noindent\textbf{Material agent part-level understanding.}
Similar to the idea of chain of thought~\cite{brown2020language}, we first guide the agent to identify what object is in the marked area, instead of directly inferring the materials.
This allows object information to be taken into account during material reasoning, enabling full use of the extensive world knowledge stored in LLMs. 
In this part of the prompt, we first explain the prompt image and ask the agent to zoom in on the corresponding area for better recognition. Additionally, we highlight some considerations, such as occlusion issues, and then ask the agent to perform object recognition step-by-step.

\noindent\textbf{Material agent object-aware material reasoning.}
Next, we ask the agent to combine pixel information and object information to reason about all possible material types and to predict results in the order of probability.
We have divided the materials in the \textit{DiffMat} library into 19 categories including asphalt, bricks, ceramics, concrete, fabric, \emph{etc}.,
and require the agent to only choose from these types. We provide the detailed output format and an example for in-context learning~\cite{min2022rethinking}.

\begin{figure}[!t]
    \centering
    \includegraphics[width=1\linewidth]{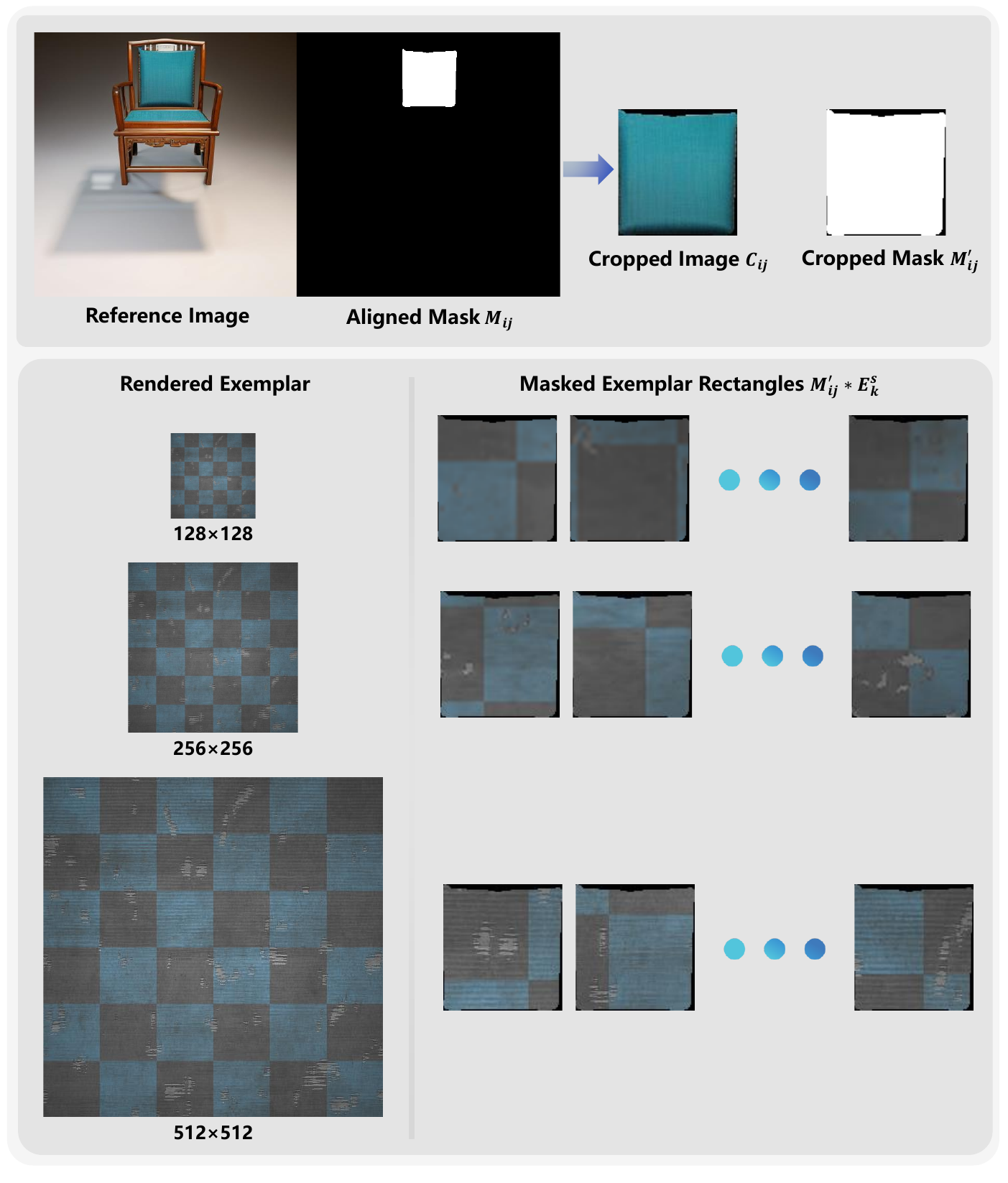}
    \caption{Material Matching Example. Above: the cropped image $C_{ij}$ and the cropped mask $M_{ij}'$ obtained through the bounding box. Below: the masked exemplar rectangles $M_{ij}'*E_k^s$ randomly sampled on the rendered exemplars at different scales of a fabric material.
    }
    \label{fig:vgg}
\end{figure}

\noindent\textbf{Material matching based on weighted VGG loss.}
Then we design a multi-scale weighted material matching method based on VGG loss~\cite{Simonyan15}, considering the order of likelihood. Specifically, we first use the mask $M_{ij}$ of material part $j$ to obtain its bounding box, the cropped mask $M_{ij}'$, and the cropped image $C_{ij}$ from the reference image $I_i$.
Next, we randomly sample a certain number of rectangles $E_k^{s}$ on the rendered exemplars at different scales ($128\times 128$, $256\times 256$, $512\times 512$) for all materials of the predicted possible types, according to the bounding box size. Then, we apply the cropped mask $M_{ij}'$ to these sampled rectangles to obtain masked exemplar rectangles, thereby mitigating the effects of scale or spatial transformations on the texture 
patterns. Then we assign the initial material $\mathcal{G}_j^{0}$ through:
\begin{multline}
\label{eq:vgg}
\mathcal{G}_j^{0}=\underset{\mathcal{G}}{\text{argmin}}\,\alpha^{O(\mathcal{G})-1}~\frac{1}{K}\sum_{s=1}^{3} \sum_{k=1}^{K} \sum_{l} \sum_{x}\\
\quad( \mathcal{F}_\mathrm{vgg}^l(C_{ij})-\mathcal{F}_\mathrm{vgg}^l(M_{ij}'*E_k^{s}))^2,
\end{multline}
where $\mathcal{G}$ is a procedural material, $\alpha^{O(\mathcal{G})-1}$ is the weight related to the likelihood order $O(\mathcal{G})$ of the type of material $\mathcal{G}$ ($O(\mathcal{G})$ starts from 1, 
$\alpha$ is set to $\frac{5}{3}$),
$K$ is the number of sampled rectangles at each scale, $s$ refers to different scales ($s=1$ is $128\times 128$, $s=2$ is $256\times 256$, $s=3$ is $512\times 512$), $\mathcal{F}_\mathrm{vgg}^l(\cdot)$ refers to the normalized VGG feature map extracted from layer $l$, $\sum_{x}$ denotes the sum over pixels $x$, and $E_k^{s}$ is the sampled rectangle on the rendered exemplar of material $\mathcal{G}$ at scale $s$. To aid intuitive understanding of Eq.~\ref{eq:vgg}, we present an example in Fig.~\ref{fig:vgg}.

\subsection{Material and Lighting Optimization}
\label{ssub:optimization}
In this section, we detail our optimization process and the optimization function.
Given the object mesh, our method adaptively creates the floor and four walls in the scene of the differentiable renderer, and adaptively places initial area lights on each wall and above the object. 
The intensity of these lights is included in the optimization. We set the light bounce count to 2 to enable global illumination. 
Our optimization process consists of two stages. In the first stage, we keep the lighting fixed and optimize only the parameters of the procedural materials. In the second stage, we jointly optimize the lighting and the procedural materials.

We denote the rendering function of the differentiable renderer PSDR-CUDA as $R(\cdot)$, and represent the parameters of all assigned procedural materials for the mesh as $\theta$. 
To obtain the $U$-$V$ maps, we use `Smart UV projection' in Blender for each material part.
Given the rendered images $R(\theta)_i$ and the reference images $I_i$, where $i$ indexes the camera view, we calculate the loss in Eq.~\ref{eq:total} between them to backpropagate the gradient from the pixel level to the computational graph of procedural materials, thereby optimizing the parameters $\theta$. 
We use Adam~\cite{kingma2014adam} as the optimizer. The total optimization objective is:
% \vspace{-0.1cm}
\begin{equation}
\mathcal{L}_{\text{total}}=\lambda_1 \mathcal{L}_{\text{resized}}+\lambda_2 \mathcal{L}_{\mathrm{\text{pixel}}}+\lambda_3 \mathcal{L}_{\mathrm{\text{stat}}}+\lambda_4 \mathcal{L}_{\mathrm{\text{gram}}},
\label{eq:total}
% \vspace{-0.1cm}
\end{equation}
where $\lambda_{1-4}$ are weights, $\mathcal{L}_{\text{resized}}$ is the $\mathcal{L}_1$ difference between the reference image and rendered image (downsized to $1/8$), $\mathcal{L}_{\text{pixel}}$ is the average $\mathcal{L}_1$ difference between the reference image and rendered image with each material part mask, $\mathcal{L}_{\text{stat}}$ is the mean absolute difference of the statistics (mean $\mu$ and variance $\sigma^2$) of the masked pixels of each material part between the two images, and $\mathcal{L}_{\text{gram}}$ is the average $\mathcal{L}_1$ difference of the Gram matrix texture descriptor~\cite{gatys2016image} $T_g$ for each part. These four sub-objectives are defined as follows:
% \vspace{-0.2cm}
\begin{equation}
\mathcal{L}_{\text{resized}}=\sum_{i=1}^{n} \mathcal{L}_1(R(\theta)_i',I_i'),
\label{eq:resize}
\end{equation}
% \vspace{-0.2cm}
\begin{equation}
\mathcal{L}_{\text{pixel}}=\frac{1}{\sum_{i=1}^{n} m_i } \sum_{i=1}^{n}\sum_{j=1}^{m_i} \mathcal{L}_1(M_{ij}*R(\theta)_i,M_{ij}*I_i),
\label{eq:pixel}
\end{equation}
% \vspace{-0.4cm}
\begin{multline}
\mathcal{L}_{\text{stat}}=\frac{1}{\sum_{i=1}^{n} m_i } \sum_{i=1}^{n}\sum_{j=1}^{m_i}\\ \left| \mu (M_{ij}*R(\theta)_i)\!-\!\mu (M_{ij}*I_i)\right|+\\
| \sigma^2 (M_{ij}*R(\theta)_i)\!-\!\sigma^2  (M_{ij}*I_i)|, 
\label{eq:stat}
\end{multline}
% \vspace{-0.6cm}
\begin{multline}
\mathcal{L}_{\text{gram}}=\frac{1}{\sum_{i=1}^{n} m_i } \sum_{i=1}^{n}\sum_{j=1}^{m_i} \\
\mathcal{L}_1(T_g(M_{ij}*R(\theta)_i),T_g(M_{ij}*I_i)),
\label{eq:gram}
\end{multline}
where $i$ represents the camera view index, $n$ represents the total number of camera views, $j$ refers to the material part index, $m_i$ refers to the total number of material parts under camera view $i$, $R(\theta)_i'$ and $I_i'$ are downsized images, and $M_{ij}$ is the mask for material part $j$.

\section{Experiments}
\label{sec:experiments}
\subsection{Experimental  Setup}
\noindent\textbf{Datasets.}
We performed experiments using the 3D-FUTURE~\cite{fu20213d}, 3DCoMPaT~\cite{li20223d_compat} and Objaverse~\cite{deitke2023objaverse} datasets. For detailed datasets description, please refer to the appendix.

\noindent\textbf{Baselines.}
We compared our method with two state of the art text-driven texture generation methods, TEXTure~\cite{richardson2023texture} and Text2Tex~\cite{chen2023text2tex}. 
\begin{figure*}[!t]
    \centering
    \includegraphics[width=1\linewidth, trim={20mm 0 20mm 17mm}, clip]{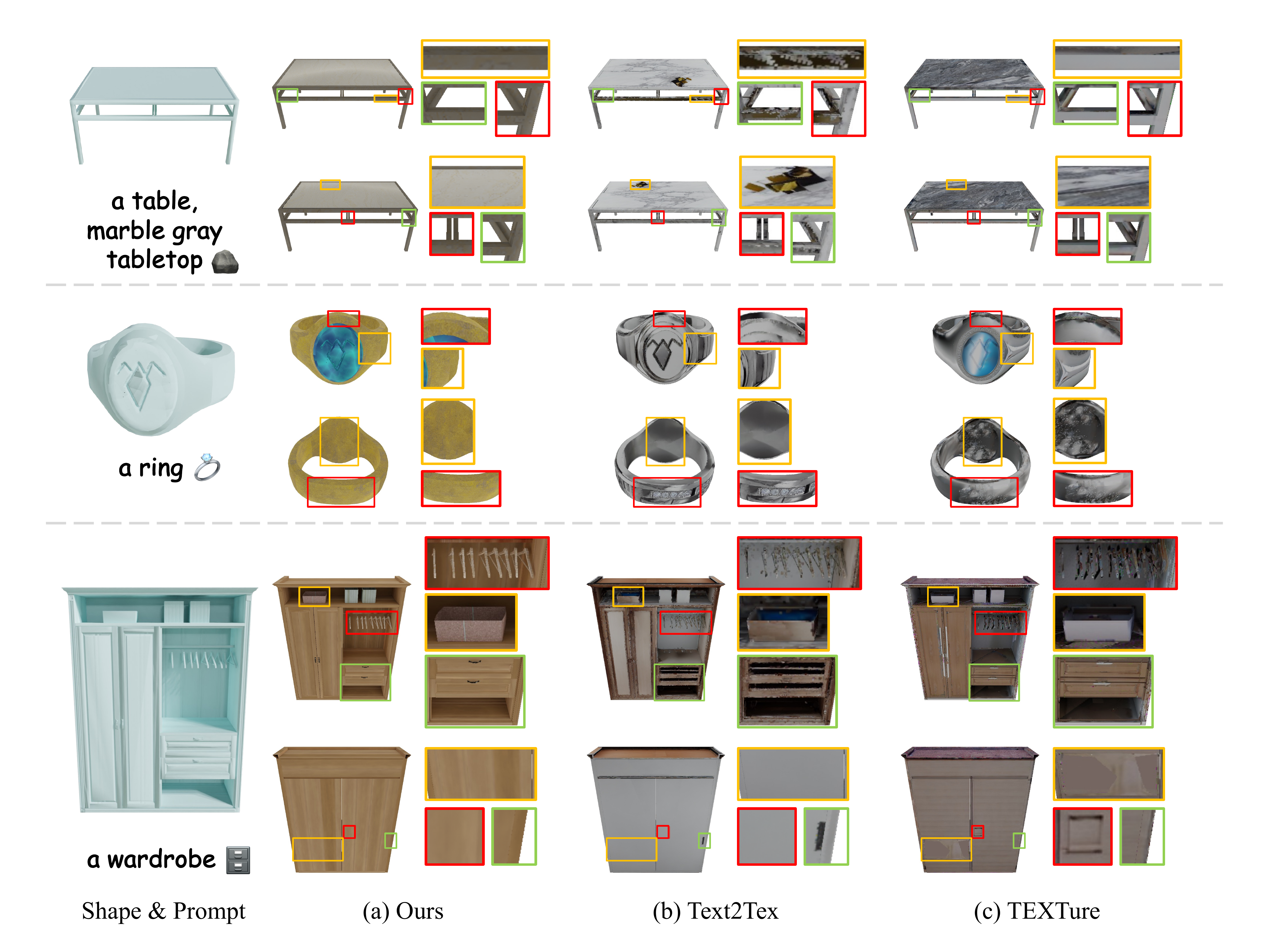}
    % \vspace{-0.4cm}
    \caption{
    Qualitative comparisons with Text2Tex and TEXTure from front and back views.
    }
    \label{fig:exp2}
    % \vspace{-0.3cm}
\end{figure*}
These two approaches utilize Stable Diffusion~\cite{rombach2022high} to generate 2D images, which are then projected onto the mesh. 
They generate complete textures through multiple iterations using an in-painting approach.
However, they only produce diffuse textures with baked lighting. As a result, they cannot support relighting, so in the relighting experiment of Sec.~\ref{ssub:relight}, we only present the results of our method.
Please refer to our supplementary video for intuitive qualitative results.

\subsection{Comparison to Baselines}
\begin{table}[t!]
\centering
\caption{
User study results on Overall Quality and Text Fidelity with 30 respondents. 
}
\label{tb:user_study}
\begin{tabular}{c c c} 
\toprule
Method & Overall Quality($\uparrow$) &  Text Fidelity($\uparrow$) \\
\midrule
Text2Tex~\cite{chen2023text2tex} & 3.44 & 3.95 \\
TEXTure~\cite{richardson2023texture} & 3.16 & 3.62 \\
Ours & \textbf{4.64} & \textbf{4.53} \\
\bottomrule
\end{tabular}
\end{table}

\begin{figure*}[!t] % p
    \centering
    \includegraphics[width=0.92\linewidth]{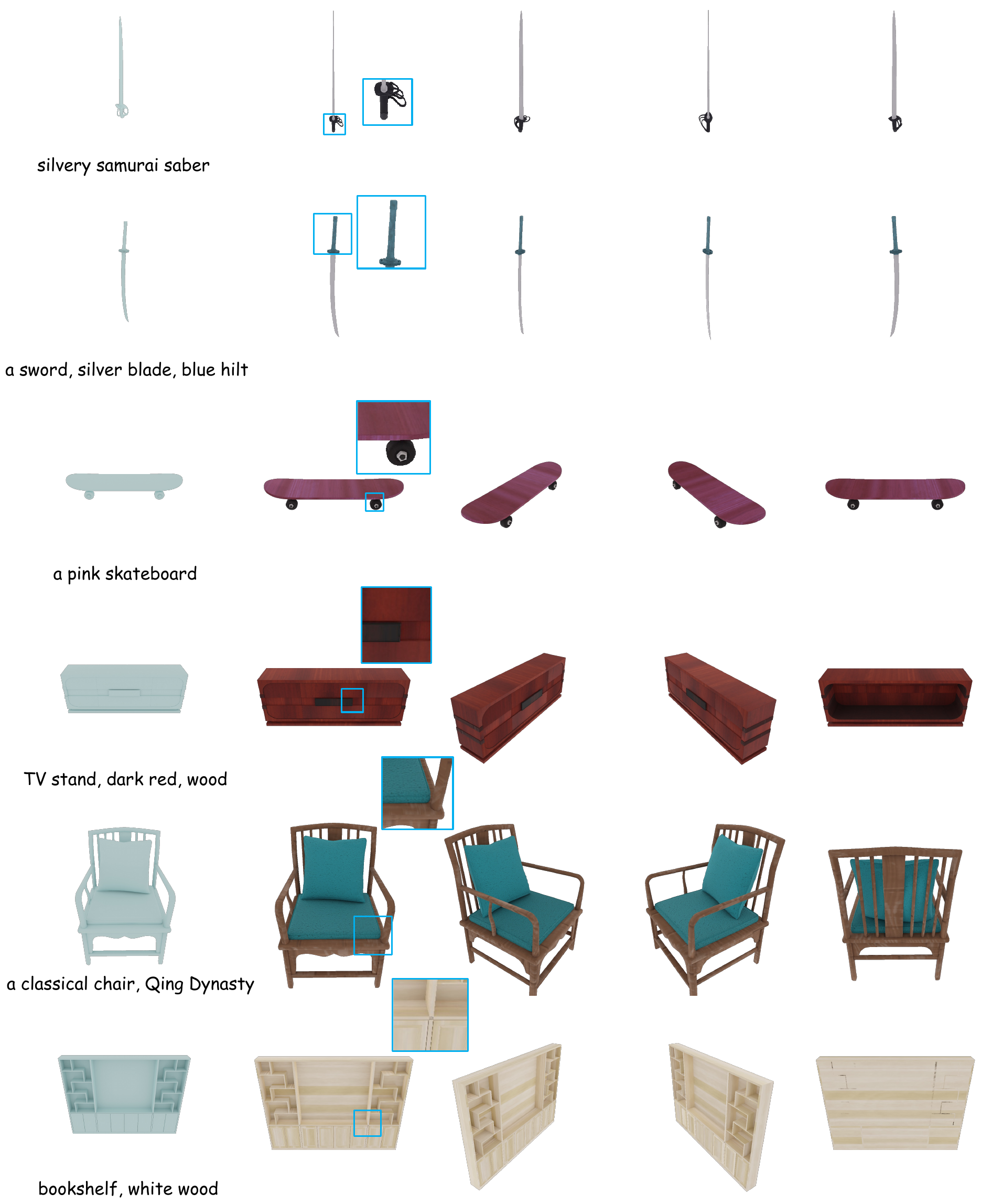}
    % \vspace{-0.2cm}
    \caption{More Texturing Results. For each sample, we present four different viewpoints.}
    \label{fig:more}
    % \vspace{-0.4cm}
\end{figure*}

This section provides qualitative and quantitative comparisons with the baselines.
Following TEXTure~\cite{richardson2023texture}, we conducted a user study with the same configuration to quantitatively evaluate the quality of text-driven texture generation.
Specifically, we prepared 10 test examples (shapes and text descriptions) for each method. We asked 30 participants to rate the overall quality and fidelity to the text, on a scale from 1 to 5. The results are shown in Table~\ref{tb:user_study}, indicating that our method achieves the best scores among all methods. 
In Fig.~\ref{fig:exp2}, we qualitatively compare the texturing results of different methods from both front and back viewpoints.
Both TEXTure and Text2Tex produce some artifacts, as both methods require projecting the generated 2D images onto the mesh based on the inpainting approach, resulting in missing textures in some obscured areas. Moreover, due to the lack of multi-view datasets training, the Stable Diffusion model utilized by these two methods can result in inconsistencies across different views, thus causing the Janus (multi-faced) problem. In contrast, our method can generate textures that maintain consistency across multiple views and produce photorealistic texturing results. More texturing results are shown in Fig.~\ref{fig:more}.

\subsection{Relighting Results}
\label{ssub:relight}
Our method can generate various maps as defined in the procedural material computation graphs, including base color, normal, roughness, \emph{etc}., which support physically based rendering, making the texturing results truly usable for downstream tasks. Thus, in this section, we present the relighting results of our method in the 3D design software Blender.

\begin{figure*}[!t]
    \centering
    \includegraphics[width=1\linewidth, trim={17mm 0 17mm 7mm}, clip]{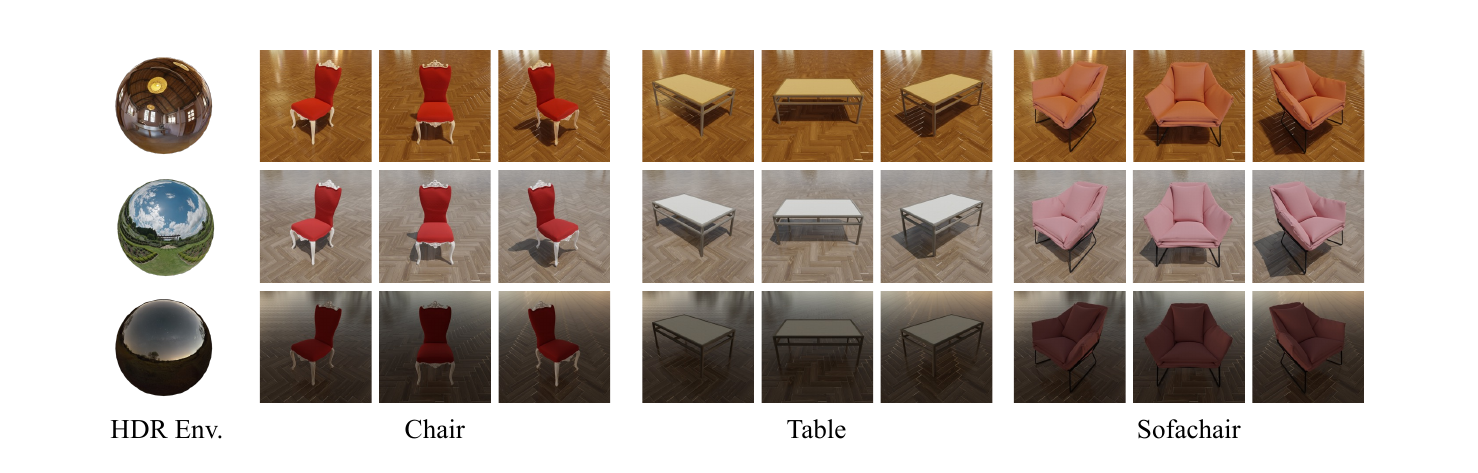}
    % \vspace{-0.4cm}
    \caption{Different Ambient Lighting. We present the relighting results for three meshes in Blender under three different ambient lighting scenarios, from three viewpoints.
    }
    \label{fig:exp1_hdr}
    % \vspace{-0.2cm}
\end{figure*}

\noindent\textbf{Different ambient lighting.}
We first use different ambient lighting scenarios (indoor warm light, daytime outdoor, nighttime outdoor) to relight our textured meshes in Fig.~\ref{fig:exp1_hdr}.

\begin{figure*}[!t]
    \centering
    \includegraphics[width=1\linewidth, trim={7mm 0 7mm 8mm}, clip]{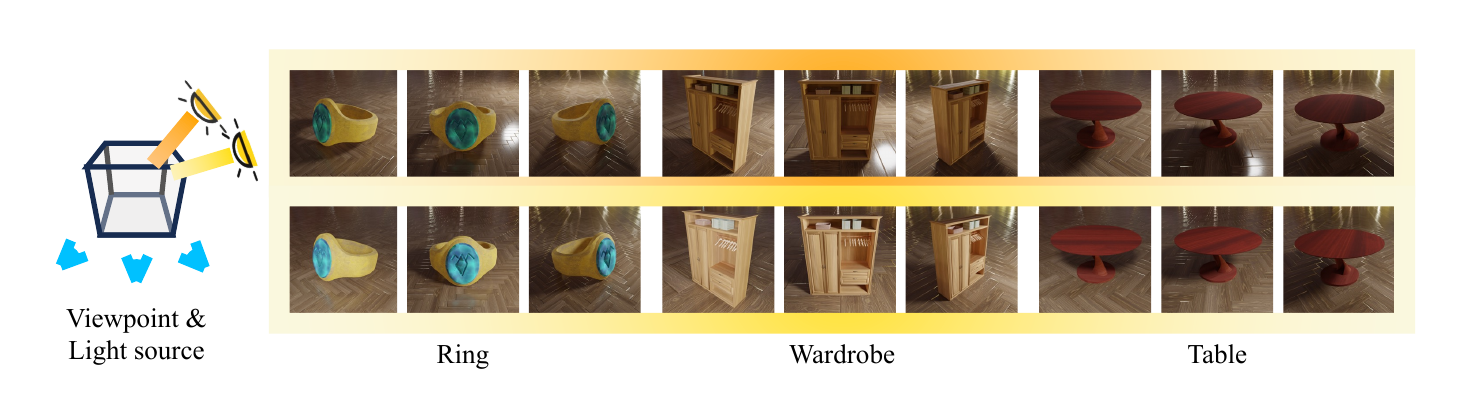}
    % \vspace{-0.5cm}
    \caption{Different Point Light Positions. 
    The top row corresponds to a point light positioned northeast, and the bottom row corresponds to southeast, each shown from three viewpoints.
    }
    \label{fig:exp1_light}
    % \vspace{-0.5cm}
\end{figure*}

\noindent\textbf{Different point light positions.}
We then keep the ambient lighting fixed, add a point light, and change its position (to the northeast and southeast, respectively) to relight the object, showcasing shadow and highlight details in Fig.~\ref{fig:exp1_light}.

\subsection{Ablation Studies}
\begin{figure}[!t]
    \centering
    \includegraphics[width=0.9\linewidth]{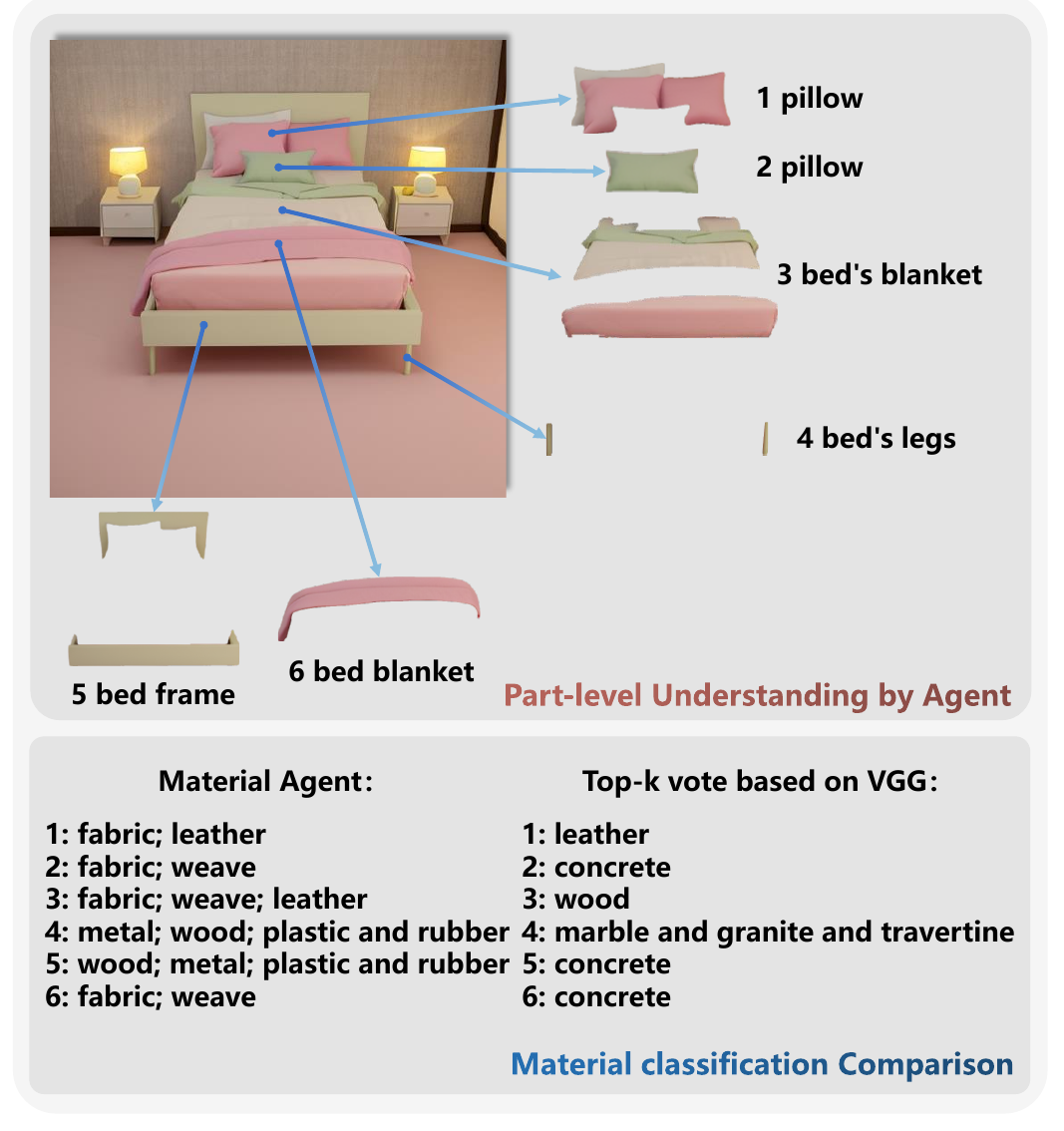}
    % \vspace{-0.2cm}
    \caption{Material Classification.
    }
    \label{fig:ablation-agent}
\end{figure}

\noindent\textbf{Material agent for material classification.}
Through the material agent, we can achieve object-aware material classification with the order of likelihood. Moreover, the number of predicted class labels is unconstrained. Here, we compare against a simple baseline. This baseline calculates the VGG loss for all materials in the library using Eq.~\ref{eq:vgg} (with $\alpha=1$), then selects the 10 materials with the lowest loss and uses voting to determine the most frequently occurring material type as the predicted type. 
We present the comparison results in Fig.~\ref{fig:ablation-agent}, where the upper part shows part-level understanding by the agent and the lower part compares the material classification for different material parts.

\begin{figure}[!t]
    \centering
    \includegraphics[width=1\linewidth]{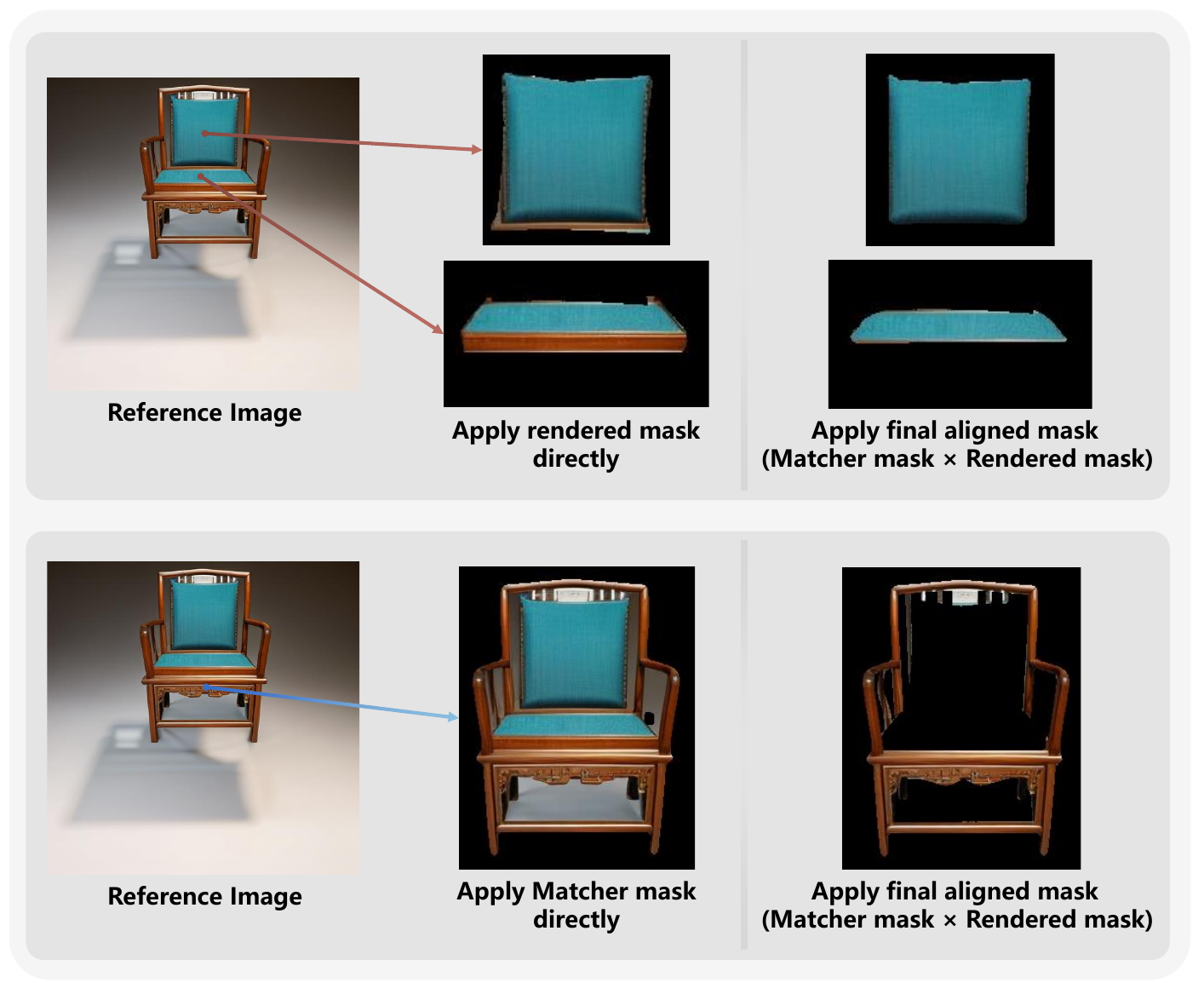}
    \caption{Final Aligned Mask. We show the three material parts of a chair.
    We cannot directly apply the rendered mask or the Matcher mask to the reference image.
    }
    \label{fig:ablation-matcher}
\end{figure}

\noindent\textbf{Final aligned masks.}
For a material part, accurately obtaining its mask on the reference image is crucial for guiding the material agent and calculating the optimization function (Eq.~\ref{eq:total}). As shown in Fig.~\ref{fig:ablation-matcher}, since the reference image is not well aligned with the mesh geometry, the rendered mask cannot be directly applied to the reference image. Thus, we obtain the final aligned mask by intersecting the Matcher~\cite{liu2024matcher} mask and rendered mask.

\section{Limitation}
Our approach also has some limitations. Since it optimizes the parameters of individual nodes within the procedural material graph, it is fundamentally constrained by the materials available in the procedural material library. As a result, it is currently incapable of generating complex textures, such as letters or characters. Nevertheless, if such textures already exist within the library, our method can still support their generation. A potential future direction could involve a two-stage generation process, where complex textures like characters are synthesized and overlaid on top of procedural materials, further expanding the expressiveness of our approach.

\section{Conclusion}
\label{sec:Conclusion}
We present a new method to generate high-fidelity materials for 3D shapes given textual prompts.
Unlike previous texture generation methods that generate RGB textures containing baked lighting, our method can generate diverse texture maps that enable physically-based rendering and relighting.
Additionally, our approach is based on procedural material modeling, which allows us to inherit many advantages of procedural materials.
Currently, our method requires pre-segmented material parts which could be improved by integrating it with shape analysis methods \cite{hanocka2019meshcnn,lahav2020meshwalker,yin2023sai3d} in the future.

\appendix
\renewcommand{\thefigure}{A\arabic{figure}}
\setcounter{figure}{0}
\section*{Appendix}
This appendix provides further details and results omitted from the main paper. Specifically, we provide additional implementation details in Sec.~\ref{sec:details} and experimental details in Sec.~\ref{sec:Evaluation_Details}. In Sec.~\ref{sec:more_results}, we present more comparisons involving our method.
Please refer to our supplementary video for more vivid method explanations and qualitative results.

\section{Implementation Details}
\label{sec:details}

\subsection{Camera Views Selection}
Given two elevation angles and the distance to the object, we use spherical sampling to uniformly sample $n$ camera coordinates for each elevation. We then construct the extrinsic camera parameters using the “Look At" approach. 
We calculate the number of pixels for each material part under each viewpoint through the rendered mask $M_R$. 
We select a sufficient number of cameras from these $2\times n$ cameras to ensure that the pixel count for each part exceeds a threshold (set to 500). To deal with the slight differences between reference images from different views, we select only one reference image for each material part.

Specifically, the front view of the larger elevation will first be selected, because, using this view, the reference image generated by Stable Diffusion~\cite{rombach2022high} will be more realistic and most material parts will exceed the pixel count threshold. 

For material parts with a pixel count below the threshold under this front view, including those unseen parts, we then adaptively select $k$ cameras based on their pixel counts.
To balance the goals of minimizing the number of selected cameras and maximizing the pixel count for each material part, we adopt three steps to select the cameras for the remaining material parts. 
The first step is to select the “unavoidable" cameras. For each material part, we compare the maximum pixel count under the $2n-1$ viewpoints with the threshold. If the maximum value is less than the threshold, we select the camera corresponding to the maximum pixel count to ensure sufficient visibility for this material part. Through this step, we select $n_1$ viewpoints. Next, we select cameras for the remaining material parts.
The second step is to select the minimal subset of cameras that satisfy the threshold condition.
We mark the cameras for each material part where the pixel count exceeds the threshold, then select the minimal subset of marked cameras to ensure that each material part meets the threshold condition. Through this step, we select $n_2$ viewpoints.
The third step involves finding the intersection of the initially selected front view, the $n_1$ viewpoints selected in the first step, and the $n_2$ viewpoints selected in the second step, and using the resulting intersection as the final set of chosen cameras. Through this step, we can obtain the final $k$ cameras.
Note that during optimization, each material part will only use the viewpoint selected for it. 

\begin{figure*}[!p]
    \centering
    \includegraphics[width=0.92\linewidth]{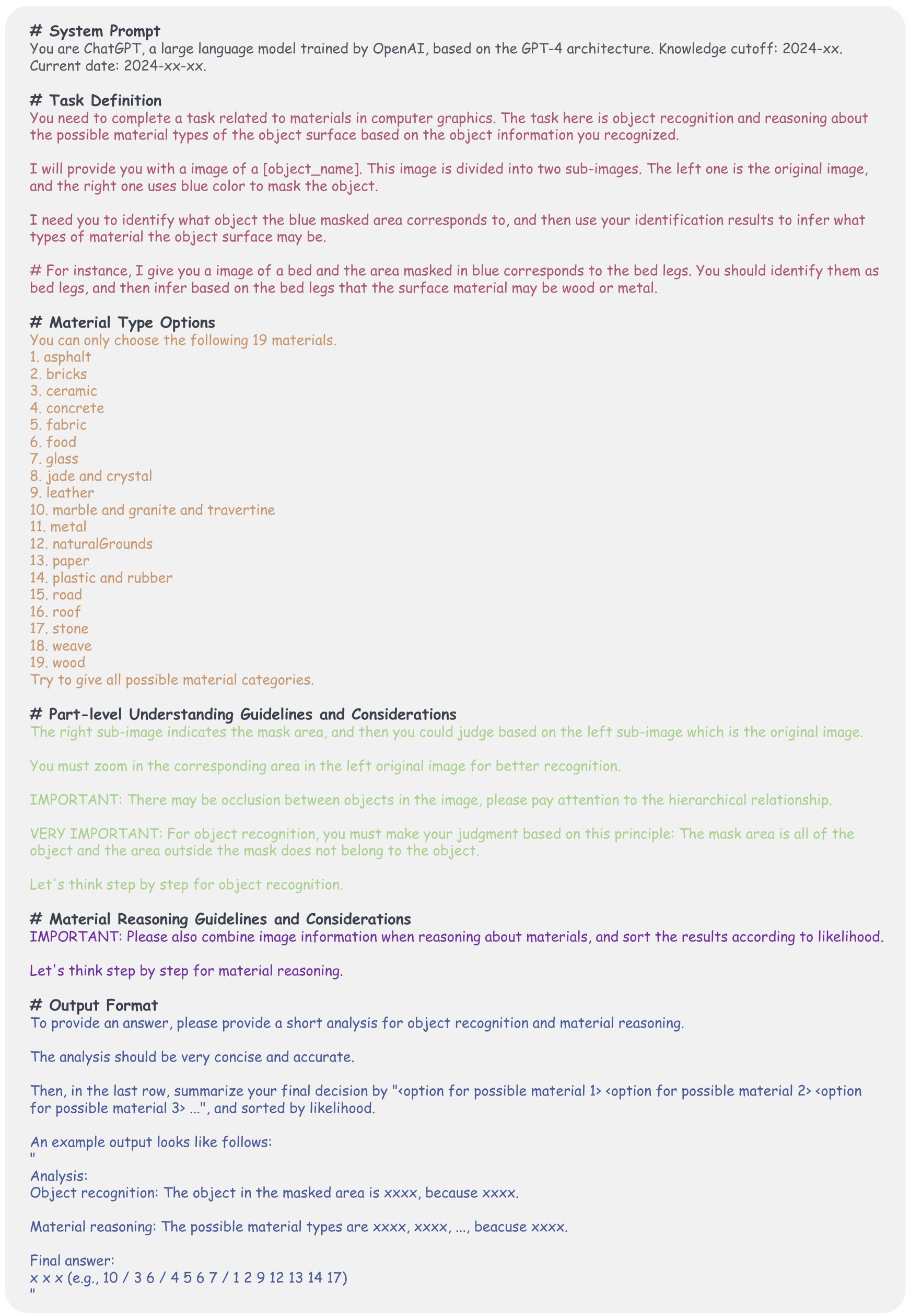}
    \caption{{Prompt template.}
    }
    \label{fig:prompt}
\end{figure*}

\subsection{Smoothing Matcher Masks}
Matcher~\cite{liu2024matcher} can segment anything by using an in-context example without training. As shown in Fig.~\ref{fig:pipeline} of the main paper, we use the rendering of the object textured by white material under specific ambient lighting and the mask $M_R$ of a material part on this rendered image as the in-context example. Then, the Matcher will output the corresponding mask $M_I$ for the reference image. 

We experimented with different ambient lighting to illuminate the objects for the in-context example. In the end, we choose the indoor uniform white light for the ambient lighting, because the segmentation of Matcher will be more accurate.

The masks output by Matcher typically contain noise. Therefore, we subsequently smooth the masks through three image processing steps. We first apply median filtering to attempt to connect regions and remove some outliers. In the second step, we use the morphological erosion operation on the binary images (masks) to remove some outlier areas. 
Finally, for each Matcher mask $M_I$, we count the number of pixels $\{P_1, \dots, P_i, \dots\}$ in each connected region $\{R_1, \dots, R_i, \dots\}$, where $P_i$ is the number of pixels of region $R_i$. We calculate the smallest number of pixels $P_r$ for all connected regions in the rendered mask $M_R$. We then set the connected regions with a pixel count smaller than $\beta P_r$ ($\beta=0.5$) to False (\ie, Masked) for every region in $\{R_1, \dots, R_i, \dots\}$.

\subsection{Material Agent Prompt Design}
\noindent\textbf{Prompt image.}
We choose the reference image generated by Stable Diffusion~\cite{rombach2022high} to construct the prompt image rather than the rendered image of the object textured by white material under specific ambient lighting, because the former carries more information, leading to more accurate part-level understanding by MLLMs.
To support part-level understanding, we tested four object labeling schemes: (i) obtaining contours of the material part through mask $M_{ij}$, then outlining the contours with red lines; (ii) building upon (i) by horizontally concatenating with the reference image; (iii) blending mask $M_{ij}$ with the reference image, overlaying the mask onto the reference image; and (iv) building upon (iii) by horizontally concatenating with the reference image.
Overall, we found that the fourth solution (\ie, the prompt image in Fig.~\ref{fig:Agent}) leads to more accurate object recognition results.

\noindent\textbf{Text prompt template.} Our prompt template is shown in Fig.~\figref{fig:prompt}.
Given the name of the object (\eg, chair), a specific prompt text can be derived from the carefully designed template by replacing \textit{[object\_name]} in the template with the name of the object. Note that all material parts share the same prompt text, rather than deriving a prompt for every material part.

\begin{figure*}[!t]
    \centering
    \includegraphics[width=1\linewidth, trim={0 0 0 0mm}, clip]{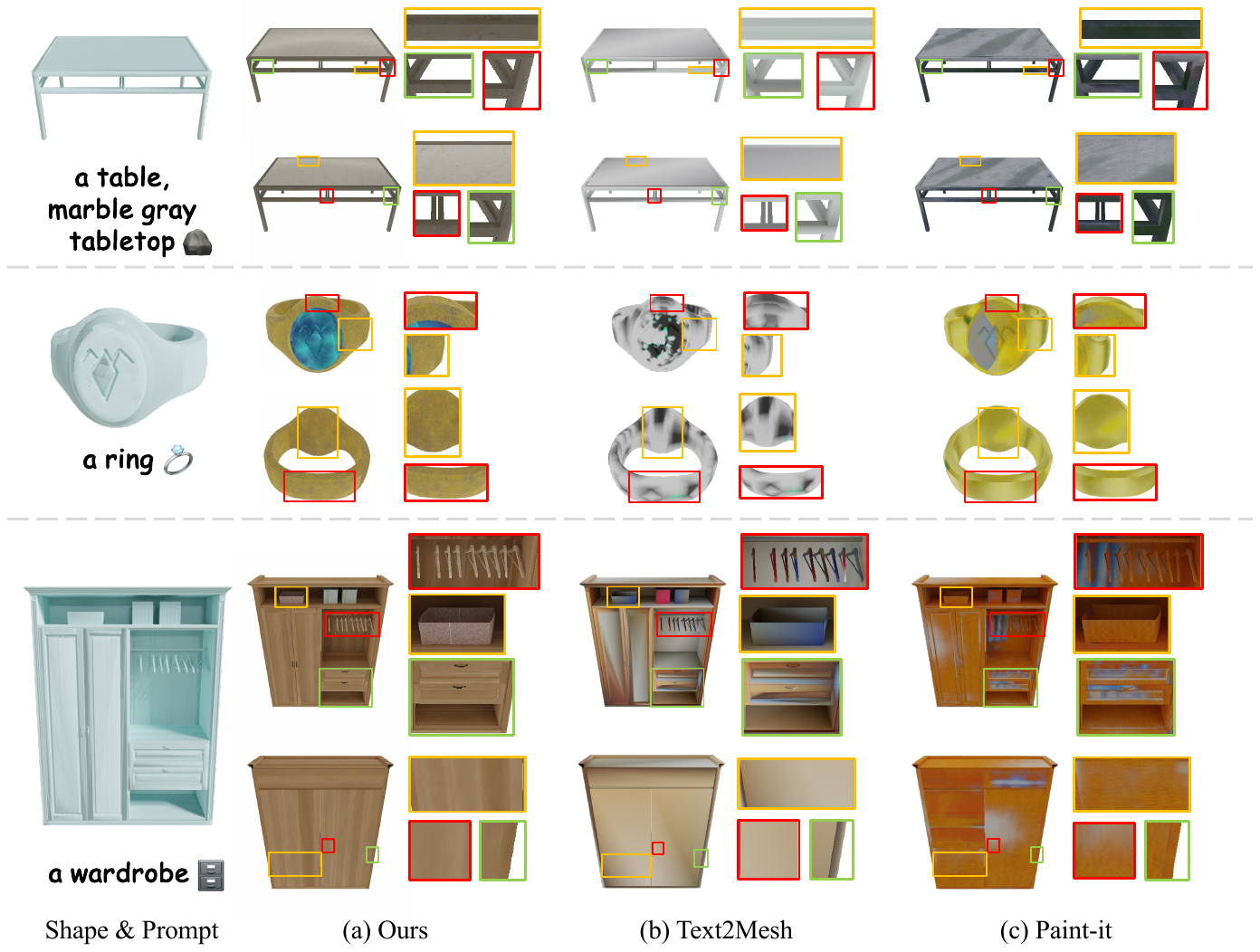}
    \caption{{Qualitative comparison to Text2Mesh and Paint-it.} It can be seen that our approach can produce more realistic and consistent textures. 
    }
    \label{fig:comparison}
    \vspace{-0.2cm}
\end{figure*}

\section{Experimental Details}
\label{sec:Evaluation_Details}
\subsection{Datasets}
We evaluated our method on three datasets including 3D-FUTURE~\cite{fu20213d}, 3DCoMPaT~\cite{li20223d_compat} and Objaverse~\cite{deitke2023objaverse} dataset.

3D-FUTURE~\cite{fu20213d} is a richly-annotated large-scale dataset of 3D furniture shapes in household scenarios. It contains 9,992 unique industrial 3D CAD meshes of furniture with related attributes such as category, style, theme, \emph{etc}., and high-resolution informative textures developed by professional designers.

3DCoMPaT~\cite{li20223d_compat} is a large-scale 3D dataset of more than 7.19 million rendered compositions of materials on parts of 7,262 unique 3D models. 3DCoMPaT covers 43 shape categories, 235 unique part names, and 167 unique material classes that can be applied to parts of 3D objects. This dataset primarily focuses on stylizing 3D shapes at part-level with compatible materials. Therefore, each object in this dataset has material part segmentation annotation.

Objaverse 1.0~\cite{deitke2023objaverse} is a large-scale corpus of high quality, richly annotated 3D objects, which contains over 800,000 3D assets designed by over 100,000 artists.
Assets in Objaverse not only belong to varied categories like animals, humans, and vehicles, but also include interiors and exteriors of large spaces that can be used, \eg, to train embodied agents.
Objaverse improves upon present day 3D repositories in terms of scale, number of categories, and in the visual diversity of instances within a category.

\subsection{Relighting in Blender}
In the relighting experiments of the main paper (\ie, Sec.~\ref{ssub:relight}), we utilized the HDRI environment maps (copyright-free) from Poly Haven~\cite{PolyHaven}, a curated public asset library for visual effects artists and game designers. For the texture maps applied to the meshes in Blender, we extracted diffuse maps, normal maps, and roughness maps from the computational graphs of the optimized procedural materials, as all computational graphs can output at least these three types of maps. Moreover, for materials classified as metal, we also employed the metallic maps.
Subsequently, we create a Blender Principled BSDF for each material part. 
The Principled BSDF in Blender is based on the OpenPBR surface shading model, and provides parameters compatible with similar PBR shaders found in other software, such as the Disney and Standard Surface models.

\section{Extended Experiments}
\label{sec:more_results}

In this section, we additionally compare our method with two other state of the art text-driven texture generation methods: Text2Mesh~\cite{Michel_2022_CVPR} and Paint-it~\cite{youwang2024paintit}. Unlike diffusion-based methods, Text2Mesh directly optimizes textures and geometries via a CLIP-based optimization function. 
In addition, the mesh texture generated by Text2Mesh is saved as per-vertex color, rather than texture maps. Similar to Text2Tex~\cite{chen2023text2tex} and TEXTure~\cite{richardson2023texture}, we removed the geometry optimization and only optimized the textures for Text2Mesh. 
Paint-it~\cite{youwang2024paintit} utilizes a neural network to represent the PBR texture maps of a mesh, and optimizes this specific neural network through a diffusion-based loss and differentiable rendering. For the texture maps generated by Paint-it~\cite{youwang2024paintit}, we constructed the PBR material nodes claimed by NVDiffRast~\cite{Laine2020diffrast} (\ie, the differentiable renderer used by Paint-it) in Blender to create the material, and then obtained the rendering results. As shown in Fig.~\figref{fig:comparison}, our method can produce more realistic and consistent textures. The texturing results generated by Text2Mesh lack texture information and show repetition because it relies solely on CLIP guidance to optimize per-vertex colors, as also demonstrated in Text2Tex~\cite{chen2023text2tex} and TEXTure~\cite{richardson2023texture}. Paint-it utilizes a randomly initialized U-Net network as a proxy for PBR texture maps and lacks material priors, leading to reduced realism in actual application scenarios (\eg, in Blender).

\section*{Declarations}
\subsection*{Availability of data and materials}
The procedural materials utilized in this study can be downloaded from the official website of DiffMat~\cite{shi2020match,li2023end} at \url{https://github.com/mit-gfx/diffmat-legacy}.

\subsection*{Author contributions}
Ziqiang Dang and Zhaopeng Cui conceived of the
presented idea, Ziqiang Dang proposed and implemented the prototype system. Ziqiang Dang, Wenqi Dong and Zesong Yang designed and performed the experiments. Ziqiang Dang and Zhaopeng Cui wrote the paper. Bangbang Yang and Liang Li contributed to the method design.
Yuewen Ma and Zhaopeng Cui supervised the project. Ziqiang Dang created the demo video. All authors discussed the results and contributed to the final manuscript.

\subsection*{Acknowledgements}
This research was partially supported by the National Natural Science Foundation of China (No.~62441222), and the Information Technology Center and State Key Lab of CAD\&CG, Zhejiang University. We also express our gratitude to the anonymous reviewers for their professional and constructive comments.

\subsection*{Declaration of competing interests}

The authors have no competing interests to declare that are relevant to the
content of this article.\\

\subsection*{Electronic Supplementary Material}
A supplementary video containing method explanations and qualitative experimental results is available in the online version of this article.

% for bibtex
% \pagebreak
%\clearpage
\bibliographystyle{CVMbib}
\bibliography{refs}

% \subsection*{Graphical abstract}

% It is a single, concise, and pictorial summary of the main findings of the article. It could either be the concluding figure from the article or better still a figure that is specially designed for the purpose, which captures the content of the article for readers at a single glance. We require it in JPG format, 300 dpi, and with the ratio of height to length $= 8 : 13$.
\end{document}